IAC-25-A3.3B.8.x96743

# Producing High-Resolution Martian Surface Temperature Maps Using VIR – TIR Relationships

Michael A. Frazer[a]*, Eriita G. Jones[a], Katarina Miljkovic[a], Gretchen Benedix[a]

[a] *Space Science and Technology Centre, School of Earth and Planetary Science, Curtin University, Kent St, Bentley, Perth, Australia 6102*
\* Corresponding Author

**Abstract**

Thermal infrared data (TIR; 8 – 15 µm) has a wide range of applications in Earth and planetary remote sensing. On Mars, this includes deriving thermal inertia (TI), which describes surface physical characteristics (e.g. particle size, degree of cementation) and is key for understanding geologic processes, assessing in-situ resource utilisation (ISRU) environments, and assisting mission planning. However, TI data from the THEMIS instrument is limited to 100 m/pixel resolution. Hyperspectral visible and near-infrared data (VIR; 0.5 – 5 µm) compliments TIR data by providing information on surface composition and is provided by the CRISM instrument at 12 m/pixel. In this work, we generate a machine learning regressor-based model to constrain relationships between THEMIS TI and CRISM VIR images at THEMIS resolution, and predict TI values from CRISM spectra with high accuracy ($R^2 \sim 0.90$, RMSE ~ 23.6 TIU). We use the model to produce a downscaled TI map at a spatial resolution of 12 m/pixel, an order of magnitude finer than currently available, revealing decametre-scale features previously unresolved in THEMIS data.

**Nomenclature**
VIR – Visible infrared; 0.4 – 1.2 µm
SWIR – Shortwave infrared; 1.2 – 4.5 µm
TIR – Thermal infrared; 4.5 – 15 µm
TIU – Thermal inertia units; $J\ m^{-2}\ K^{-1}\ s^{-1/2}$

**Acronyms/Abbreviations**
CRISM – Compact Reconnaissance Imaging Spectrometer for Mars
THEMIS – Thermal Emission Imaging System
CTX – Context Camera
MRO – Mars Reconnaissance Orbiter
MTRDR – Map-projected Targeted Reduced Data Record

## 1. Introduction

Satellite-derived remote sensing observations date back to the mid-1900s, with the Luna 3 spacecraft returning the first images of the dark side of the moon during a flyby in 1959, and TIROS-1 returning the first footage of Earth from space a year later (Fig. 1) [1]. As of 2025, there are over 300 Earth Observation (EO) satellites currently in orbit [2], and dozens more spacecraft throughout the wider the solar system.

Since then, instruments on board have improved dramatically, with higher resolution sensors, new techniques and increased temporal coverage [1]. Some trade-offs do remain, however, with different modes of remote sensing (e.g. visible, infrared, hyperspectral, LiDAR, SAR) each providing different spatial and temporal resolutions. This has driven the development of data fusion in remote sensing, in which two unique, separately acquired data sets of the same physical environment are synthesised into one which is 'greater than the sum of its parts'.

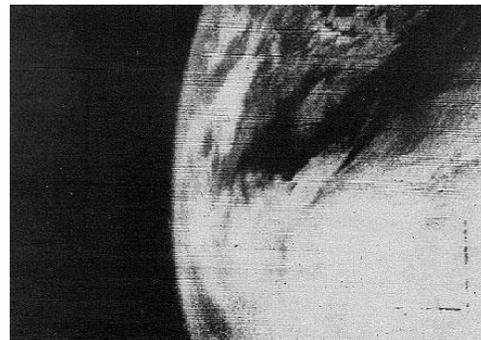

*Figure 1: First television picture from space, taken April 1, 1960 by the TIROS-I satellite. Credit: NASA.*

Data fusion has a range of applications in the EO space [3]. In this work, we employ data fusion to combine high-resolution VIR-SWIR hyperspectral data with medium-resolution TIR-derived data from separate instruments in orbit around Mars.

Section *1.1.* introduces thermal data, *1.2.* details the applications and calculation of thermal inertia, *1.3.* describes the limitations of thermal data in terms of spatial resolution, and *1.4.* is a summary of data fusion and downscaling in the EO and planetary science space. Section *2.* describes the data that we use, and section *3.* details the regression model that we train and apply on the data. Sections *4.* and *5.* detail our results and discussions respectively.

*1.1. Thermal data*

Thermal infrared (TIR) data is collected between 8 – 15 μm, and multi- and hyperspectral TIR data has a range of applications including mapping surface vegetation and minerals [4, 5]. TIR data can also be used to determine the Land Surface Temperature (LST) [6], applications of which include constraining soil moisture and evapotranspiration in agricultural contexts [7, 8], linking surface energy balance to climate changes [9], and measuring the effects of urban heat islands in populated areas [10].

On Mars, LST is used for understanding latitudinal and regional variations in the climate, measuring the presence of surface and sub-surface ice, and environmental evolution [11]. In addition to these, variations in the LST, in conjunction with albedo, are used to determine the thermal inertia of the surface, which is driven by the physical properties (grain size, cementation, etc.) of the surface material [12].

*1.2. Thermal inertia*

On Mars, a range of geologic processes have shaped the planet's surface to what it is today. Processes include (but are not limited to) fluvial, glacial/peri-glacial, and aeolian transport and erosion, lacustrine sedimentation, volcanic transport and deposition, and impact-derived fragmentation, transport and deposition. Together, these processes result in a range of informative and sometimes diagnostic physical distributions in the form of surface and shallow-subsurface permeabilities, porosities, particle sizes, sorting, layering, and cementation [13, 14]. These distributions can be constrained by measuring the thermal inertia of the surface.

The thermal inertia (TI; Eq. 1) of a region describes how it resists external temperature forcings. High TI objects resist changes in their environment, while low TI objects match their external environments more rapidly. Thermal inertia is defined as:

$$TI = \sqrt{\kappa \rho c} \quad (1)$$

where $\kappa$ is the thermal conductivity, $\rho$ is the bulk density and $c$ is the specific heat [12]. TI has units of J m$^{-2}$ s$^{-1/2}$ K, generally referred to as thermal inertia units (TIU). Variations in the bulk density and specific heat of martian surface materials are relatively small compared to variations in thermal conductivity, so TI data is largely indicative of the latter, which is in turn driven by the grain size and degree of cementation of the material [13, 15]. Table 1. shows general TI-particle size relationships.

*Table 1. Thermal inertia values for various surface materials on Mars (modified from [16]).*

| Material | Particle size | TI values (TIU) |
|---|---|---|
| Dust | 2 – 60 μm | 20 – 150 |
| Sand | 60 – 2000 μm | 150 – 400 |
| Granules | 2 – 4 mm | 400 – 800 |
| Pebbles | 4 – 250 mm | 800 – 2000 |
| Boulders/ cemented soil | > 250 mm | 2000 |

While daytime thermal infrared observations are primarily driven by the reflection of solar illumination (recording reflectance, albedo and topography), nighttime observations are dominated by emission from the surface, which is in turn driven by TI. After sunset, low TI objects cool quickly and appear dark in nighttime LST images, while high TI materials retain their heat and appear bright. For example, Fig. 2 shows a global TI map of Mars, showing variation between dust-dominated (blue, low TI) and rock- or duricrust-dominated (red, high TI) regions.

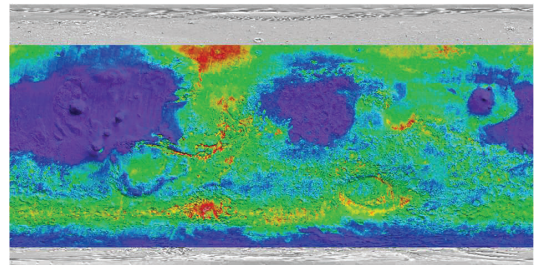

*Figure 2: Global thermal inertia map from the TES instrument. Red regions are high TI (e.g. exposed rock or duricrust), blue are low TI (fine or dust-mantled areas). Credit: NASA/Arizona State University.*

In most cases, pixel values in TI images represent some mixture of these materials – for example, an outcrop (2000 TIU) surrounded by sand (300 TIU) and covered in thin dust (100 TIU) might return a combined TI value of 800 TIU. Constraining the effects of subpixel mixing is key for decoding the returned TI values in images and understanding local surface environments.

*1.3. Limitations of thermal data*

A primary limitation of satellite-based LST/TI measurements is that TIR data is collected at coarser spatial resolutions than VIR data. For example, the multispectral EO ASTER instrument has spatial resolutions of 15, 30 and 90 m/pixel in its VIR, SWIR and TIR bands respectively (Fig. 3) [17, 18], and the Mars-based CRISM (VIR-SWIR) and THEMIS (TIR)

IAC-25-A3.3B.8.x96743

instruments provide data down to 12 and 100 m/pixel respectively [19, 20].

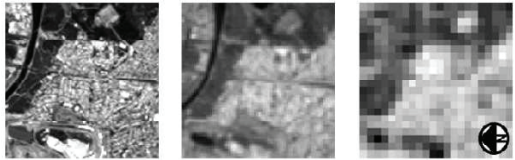

*Figure 3: Comparison of the spatial resolutions of ASTER's VNIR (15 m/pixel; left), SWIR (30 m/pixel; centre) and TIR (90 m/pixel; right) bands. From [18].*

This severely limits multi-wavelength studies, particularly of surface features ~100 m in size, which can be resolved in the VIR-SWIR but not the TIR. To address this, data fusion techniques have been developed in the EO space which link the two unique wavelength domains.

*1.4. Data fusion techniques*

The first generation of data fusion algorithms find simple linear relationships between VIR and TIR data. For example, the DisTrad [21, 22] and TsHARP [23, 24] models use linear regression to quantify the physical relationships between the VIR-derived Normalised Difference Vegetation Index (NDVI) and TIR-derived LST to estimate LST from the higher-resolution NDVI observations, and thus generate high-resolution LST maps. While these single-predictor models work well in agricultural areas where the surface is homogenous over fairly large regions (e.g. fields), they were found to struggle in areas with more spatially-dense variation (in composition and LST), and in areas where vegetation is not the driving factor in the temperature, including sparsely vegetated and/or urban environments [25].

Polynomial regressions which utilise multiple predictors (including albedo and other spectral indices) have also been employed [26-28], as well as machine-learning based regressions which use multiple spectral indices and/or the entire hyperspectral dataset [29]. They are generally less interpretable, but show better performance in more complex conditions.

In contrast to the EO-space, there has very little work in downscaling of planetary remote sensing data. Yang et al. [30] successfully downscaled Lunar LST by combining high resolution surface topography with a physical model of solar irradiance. He et al. and Christian et al. [31, 32] compared CRISM spectra to a catalogue of lab-derived spectra at a range of fixed temperatures and mineralogic compositions. They were able to separate out the contributions of composition (spectra) and temperature (single value) from each observed spectra, and thus generate CRISM-based LST and TI maps.

This work presents a new investigation into planetary-based downscaling, developing a technique inspired by previous EO work.

**2. Data**
*2.1. Local Region*

We apply our method using CRISM and THEMIS images of the north-west flank of Aeolis Mons within the ~150 km-wide Gale Crater (Fig. 4). This region shows a diversity of original and altered material as well as substantial changes in surface material, from outcrops to sand-filled pits.

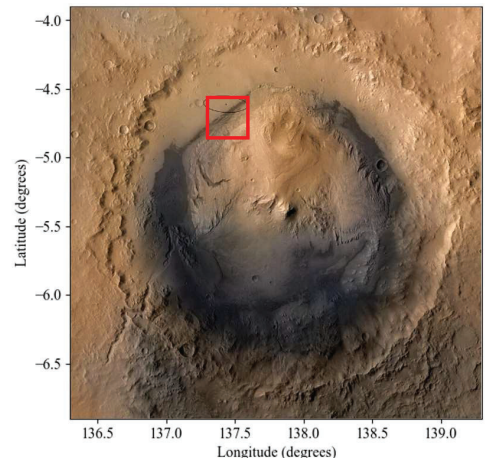

*Figure 4: Colour image of Gale crater, 154 km wide. The red box contains the area for this study, shown in Fig. 10. Credit: NASA/JPL.*

We chose this location based on several factors, including:
- the availability of an along-track oversampled (ATO; 12 m/pixel) product, and general high concentration of CRISM products available,
- the presence of a range of surface features, including distinct sand-filled pits which drive strong variation in the TI imagery,
- the presence of the Mars Science Laboratory (MSL) rover *Curiosity* within the images, whose Ground Temperature Sensor (GTS) can be used for ground-truthing, and
- the work done by [31, 32] in the same region, which can be used for comparison.

*2.2. CRISM Data*

CRISM is a hyperspectral imager onboard the MRO spacecraft [19]. The highest order science data generated by CRISM are MTRDR products, which have undergone emissivity and atmospheric corrections and been georeferenced to the surface. MTRDRs have a nominal spatial resolution of 16

IAC-25-A3.3B.8.x96743

m/pixel, with ATO achieving 12 m/pixel, and contain data across the spectral range of 362 – 3920 nm at 6.55 nm/band (producing 544 individual bands). MTRDRs have some data and bands manually removed by the CRISM team based on poor data quality.

MTRDRs are available in two forms – intensity/flux (IF) products, which contains values for each of the 544 bands, and spectral summary (SU) products, which are derived from the IF product and contain 60 spectral features (band depth, reflectance, etc.) that relate to specific features of interest (e.g. dust coverage, phyllosilicates, or ice) [33].

We use CRISM MTRDR ATO products frt00021c92_07_if164j_mtr3 (IF data) and frt00021c92_07_su164j_mtr3 (SU data). This data was collected in December 2011 at a Local True Solar Time (LTST) of 14.78 and solar longitude (Ls) of 40.58.

In addition to the bands dropped by the CRISM team, we also investigate this data for bad bands and remove several which are missing any data. This includes two SU bands ('RPEAK1' and 'BD3200'), as well as 98 IF bands, mostly between 2.8 – 3.7 μm.

*2.3. THEMIS Data*

THEMIS is a multispectral TIR imager onboard Mars Odyssey [20]. It observes across nine bands from 6 – 15 μm and has a spatial resolution of 100 m/pixel.

We use georeferenced 32-bit quantitative TI mosaic maps available on the USGS Astrogeology Science Centre website, derived from Band 9 (at 12.57 μm) of the THEMIS data following the process in Fergason et al. [13]. These data have an overall accuracy of 20%. These mosaics each cover 30° latitude by 60° longitude, and we use THEMIS_TI_Mosaic_Quant_30S120E_100mpp.

We also use CTX imagery [34], specifically from the global mosaic of Mars [35].

*2.4. Processing*

Spatial offset between the images was corrected using the shift tool in ArcGIS Pro by aligning each to the CTX imagery. Any minor misalignment led to a substantial reduction in the performance of the regressor, so we aimed to be as accurate as possible. However, due to the varying spatial resolutions, we estimate that an error of spatial misalignment on the order of ± 20 m remains in our resulting product.

We trimmed the THEMIS image to match the CRISM footprint. The resulting CRISM and THEMIS images are displayed in Fig. 5.

**3. Methods**

This work was completed in two steps – training the model at THEMIS resolution, and applying the model at CRISM resolution (Fig. 6).

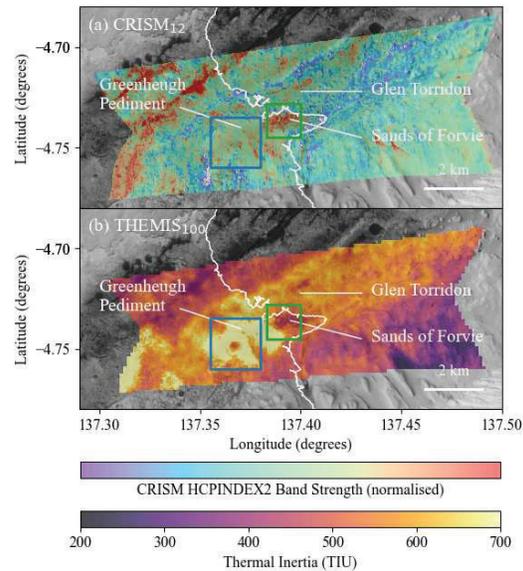

*Figure 5: (a) CRISM HCPINDEX2 and (b) THEMIS$_{100}$ TI images overlaid on a CTX background. The CRISM image shows the HCPINDEX2 band, with high (red) regions corresponding with high-Ca pyroxene abundance (basaltic sands) [32, 33]. High-TI (bright yellow) regions in the THEMIS$_{100}$ TI image correspond with exposed rock, while low-TI (dark purple) correspond with sands. The white line is the traverse of the MSL Curiosity rover, and the blue and green sections show the regions in Fig. 11.*

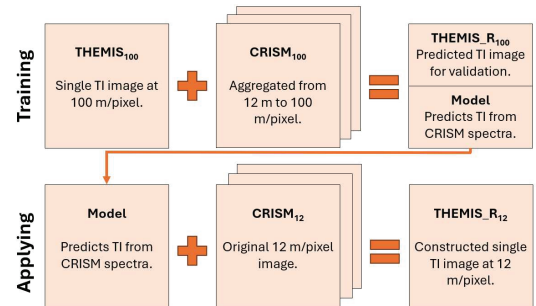

*Figure 6: Flowchart of training and applying the model.*

*3.1. Training method*

We first downsampled the original 12 m/pixel CRISM image (CRISM$_{12}$) to 100 m/pixel (CRISM$_{100}$) by averaging the ~64 finer resolution pixels within the single larger pixel. This resulted in each CRISM spectra having a single corresponding TI measurement, generating ~4485 spectrum-TI pairs. These were split randomly into training:testing populations at 75:25, resulting in 3333 training pairs and 1512 testing pairs. We used lazypredict's LazyRegressor package, which iterated over 40+ regressors (including Histogram Gradient Boosting, Random Forest, Linear

Regressor, Ridge, Lasso, etc.) to find the combination of spectral parameters which could best reproduce the TI value. Performance was assessed by calculating the coefficient of determination ($R^2$; Eq. 2) and Root Mean Squared Error (RMSE; Eq. 3) across the test data points. $R^2$ indicates the proportion of the variance explained by the model, and RMSE measures the average difference between predicted and observed values in TIU. We found that an Extra Trees Regressor demonstrated the best performance at no significant computational expense compared to the other regressors, so used that to train our final model (using scikit-learn's `ExtraTreesRegressor`). We also ran a grid search using scikitlearn's `GridSearchCV` to tune the model's parameters and maximise performance (e.g. number of estimators, $n$).

$$R^2 = 1 - \frac{\sum_{i=1}^{n}(\text{THEMIS}_i - \text{THEMIS\_R}_i)^2}{\sum_{i=1}^{n}(\text{THEMIS}_i - \overline{\text{THEMIS}})^2} \quad (2)$$

$$\text{RMSE} = \sqrt{\frac{1}{n}\sum_{i=1}^{n}(THEMIS\_R_i - \text{THEMIS}_i)^2} \quad (3)$$

We tested this process across a range of parameters using only IF, only SU, and combined IF and SU data. We found very little variation ($\Delta R^2 < 0.01$) between models using IF, SU and IF+SU, suggesting most of the complexity of the IF data is captured by the SU indices. We used only the SU indices moving forward.

The set of parameters which returns the best fit on the training data ($R^2 \sim 0.90$) also reproduces the training data exactly ($R^2 = 1.0$). This suggests overfitting, likely due to the high number of variables (58) compared to training pixels (~3500). This issue could be addressed by training on more images, however that would introduce regional variations which are beyond the scope of this investigation. Since we are not interested in the interpretability of the model or how it could be applied to other datasets, but rather its ability to recreate test data (i.e. maximise the test $R^2$) in this specific region, we allowed the model the freedom to overfit and maximise the test $R^2$. We stress that this exact model is only appropriate for this image pair, and should not be applied to other images (although a new model could be trained).

We applied this model to the $CRISM_{100}$ data to generate a reconstructed THEMIS TI image at 100 m/pixel ($THEMIS\_R_{100}$) that was used for validation and measuring the model performance (Fig. 7).

*3.2. Training validation*
We assessed the performance of the model by measuring the $R^2$ and RMSE of the original and

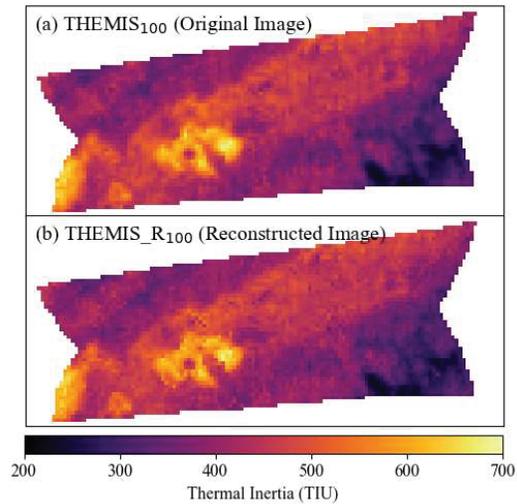

*Figure 7: (a) The original THEMIS TI image ($THEMIS_{100}$) and (b) the reproduced version ($THEMIS\_R_{100}$), composed of both testing and training pixels. The 'speckle' in $THEMIS\_R_{100}$ are the test pixels amongst training pixels. Note that the regions with the lowest and highest TI values in the original image are slightly underfitted in the reconstructed version, but that most other detail is captured reasonably well. Fig. 5 shows the $R^2$ performance of the model by comparing the test pixels between $THEMIS_{100}$ and $THEMIS\_R_{100}$.*

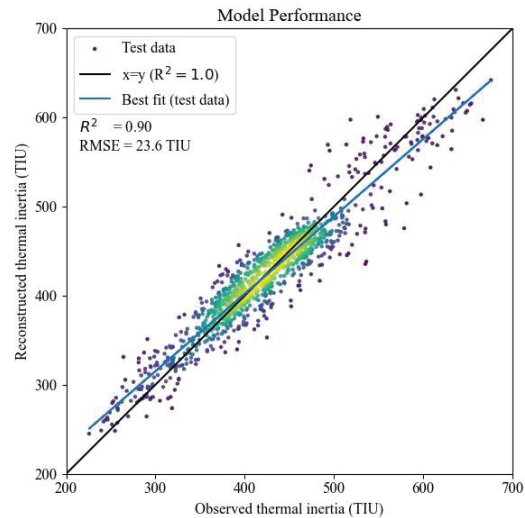

*Figure 8: A scatter plot of the 1152 'test' pixels' observed and reconstructed TI (extracted from Fig. 8). The colour represents the density of pixels in that region. The model returns a fit of R2 = 0.85 and RMSE of 29.4 TIU, suggesting a good fit.*

reproduced test pixels (Fig. 8). The model achieved an $R^2$ score of 0.90 and RMSE of 23.6 TIU on the test data,

IAC-25-A3.3B.8.x96743

indicating strong performance. Notably, the RMSE of ~30 TIU is less than the inherent 20% (~100 TIU) uncertainty in original TI values when derived from THEMIS imagery [13]. The model performed worse towards the upper edge of the dynamic range (> 500 TIU), with some pixels having offsets of ~100 TIU. This effect is also slightly evident in the original and reconstructed images in Fig. 9.

*3.3. Model application*

We then applied this model to the native high-resolution $CRISM_{12}$ data to produce a high-resolution TI map ($THEMIS\_R_{12}$).

To investigate whether the downscaled image is an accurate solution, we resampled $THEMIS\_R_{12}$ at 100 m/pixel and compared it to the original $THEMIS_{100}$ data. We observed residuals between the original and resampled images (RMSE ~ 23.6 TIU), suggesting the downscaled image is not quite an explanation of the observed image.

To 'force' consistency between the two, we take these residuals at the coarse resolution, resample them down to the finer resolution using cubic interpolation (effectively smoothing them), and apply them across $THEMIS\_R_{12}$. This enforces conservation of energy and forms the final version of $THEMIS\_R_{12}$.

**4. Results**

Fig. 9 shows the original THEMIS TI image ($THEMIS_{100}$), the newly produced downscaled image ($THEMIS\_R_{12}$), $THEMIS\_R_{12}$ re-sampled at 100 m/pixel and the subsequent residuals between it and $THEMIS_{100}$. The residuals are very low (standard deviation ~ 3.56 TIU) and contain minimal structure. The downscaled image reproduces the overall structure of the original image very well. Additional small-scale structure is visible in some regions, particularly small low-TI regions within the otherwise bright Greenheugh pediment. Section 5 investigates whether these correspond with real surface features.

**5. Discussion**

In this section we focus on specific regions within the image to consider the benefits achieved from the downscaling process. We do this with the additional context of the traverse of the MSL *Curiosity* rover, which investigated the region between sols ~1500 – 3800 (October 2016 – April 2023) (Fig. 10).

Fig. 11 focuses on specific regions which contain decametre-scale features visible in the CTX imagery that are too small to be resolved in the original $THEMIS_{100}$ imagery. For example, two rock outcrops in the northern region of the Sands of Forvie which are too small to be resolved in $THEMIS_{100}$ are visible in THEMIS $R_{12}$ and return TI values of ~ 500 TIU compared to the surrounding sand's ~ 200 TIU. A sand-filled crater (TI ~ 300 TIU) in the bottom-left of the image is also revealed, and the contours of the rock-sand interface are captured well. A small offset is present between the downscaled image and the background CTX image, with features appearing ~ 1 m pixel (10 m) further north in the downscaled

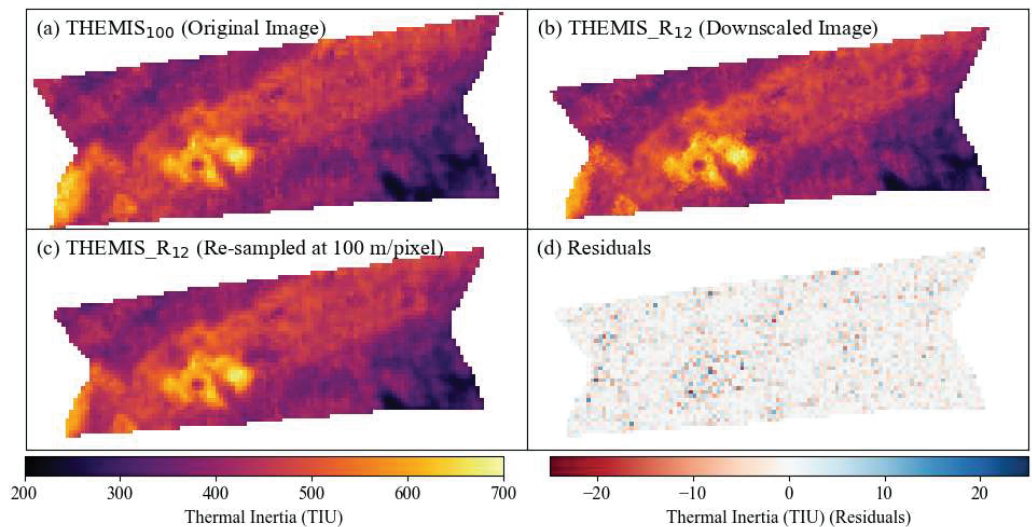

*Figure 9: (a) the original THEMIS TI image (THEMIS100), (b) the newly produced downscaled image (THEMIS_R12), (c) THEMIS_R12 re-sampled at 100 m/pixel and (d) the subsequent difference between (c) and (a) THEMIS100. The residuals are very low (standard deviation ~ 3.56 TIU) and contain minimal structure. Some additional structure is visible in (b), which is a combination of noise and real surface features being resolved (see Fig. 12).*

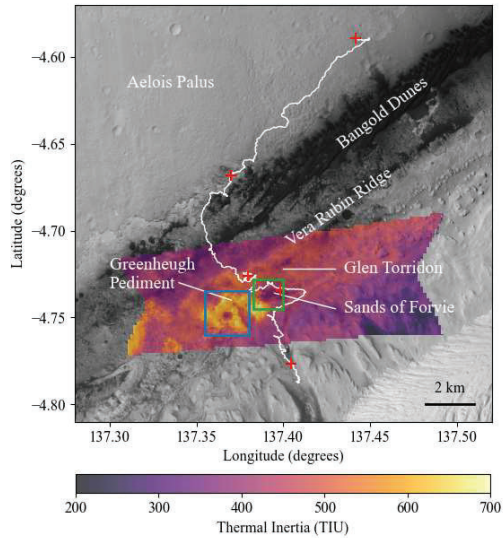

Figure 10: THEMIS_$R_{12}$ overlaid on a CTX background of the north-west flank of Aeolis Mons within Gale Crater. The white line is the traverse of the MSL Curiosity rover. Red crosses denote every 1000th sol. Note the high TI of the Greenheugh pediment and low TI of the Sands of Forvie. The blue and green boxes correspond to the zoomed regions in Fig. 8. North is up.

THEMIS image relative to the background CTX image. While visually problematic, the CTX data was not used in the downscaling process, and so did not affect model performance. However, this does highlight a limitation of using data from multiple sensors without exact precision.

Similar features are visible in the second region, in (d) and (e) – the edge of the central sand patch is more neatly resolved, and at least four smaller (< 100 m diameter) impact craters are revealed (~ 300 m to the NW, NNE and NE, and ~ 50 m SW of the central sand). TI values in these craters are around 400 – 500 TIU, which suggests some mixing of sand and coarse sand/small pebbles. The edge of the low-TI angular feature in the right of the image is also resolved more cleanly.

While not performed here, validation of this work against ground truth data is also important. The Ground Temperature Sensor (GTS) onboard *Curiosity* is able to measure the TI of ~ 10 x 10 m regions (i.e. single pixels) of the regions investigated here, and there is also room for an interesting comparative study between this work and that of He et al. and Christian et al. [31, 32]. Fig. 12 shows *Curiosity*'s view across the Sands of Forvie.

Other future work could involve testing a wider range of regressors and neural networks, as well as

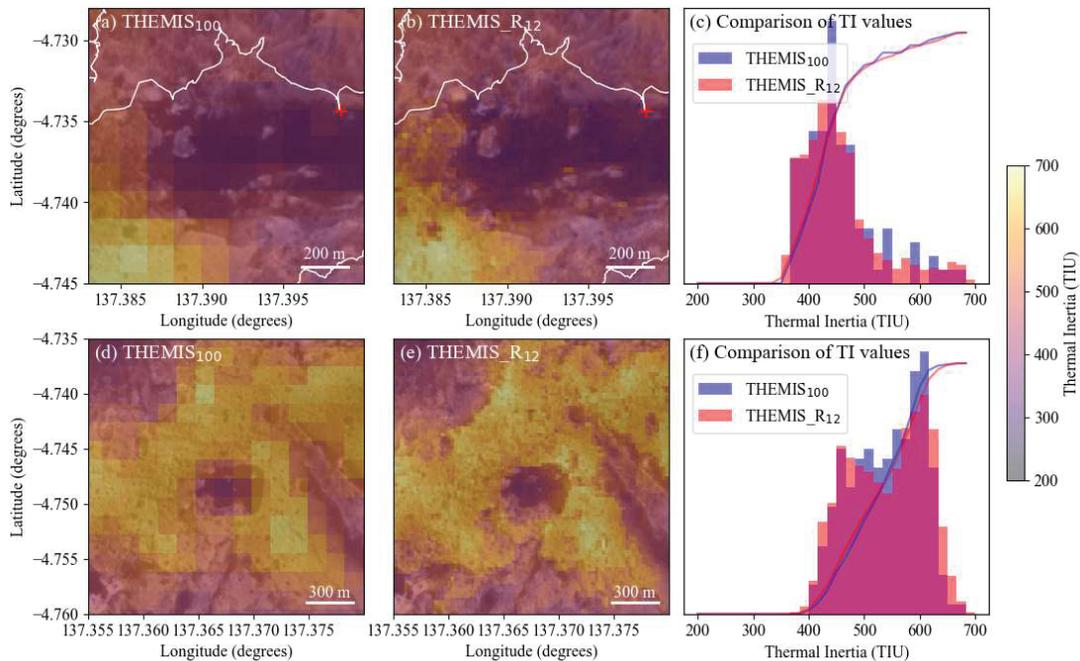

Figure 11: (a) - (c) correspond to the green region in Figs. 5 and 10, and (d) - (f) correspond to the blue region. (b) and (e) reveal small (< 100 m) features in the TI imagery which were not resolved in (a) and (d), including ~ 100 m-wide impact craters and sand-rock interfaces. The white line indicates the MSL Curiosity rover traverse, and the red cross shows the location of the rover when the image in Fig. 12 was taken.

IAC-25-A3.3B.8.x96743

deploying this tool on other CRISM-THEMIS image pairs both within the Glen Torridon region and elsewhere on Mars.

Based on previous work in the EO space, we predict that a 'one size fits all' model is likely not appropriate for use across diverse regions on Mars, and would likely require localised re-training. However, the extent of this would be interesting to explore (i.e. what surface features would require substantial retraining, and what would not?). There is the potential for a more complex model to include these regional differences in a single deployable model.

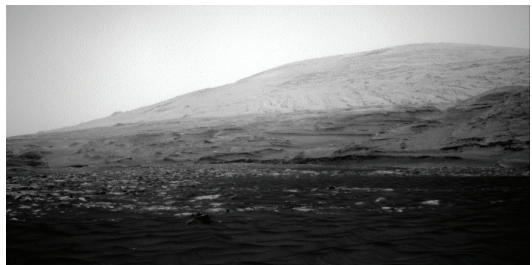

*Figure 12: MSL* Curiosity's *view across the Sands of Forvie, taken by the Right Navigation Camera on Sol 2990 at the location indicated in Fig. 11. Note the dark sand with some rock inclusions ~ 1 m wide. Credit: NASA/JPL.*

## 6. Conclusions

This work demonstrates the development and application of traditionally EO-based downscaling techniques in a planetary context to generate TI imagery at spatial resolutions almost an order of magnitude better than native resolutions. We find a good fit ($R^2 \sim 0.90$, RMSE ~ 23.6 K) when training the model for one CRISM-THEMIS image pair in the Glen Torridon region, and the subsequent downscaled image a) is consistent with the original data and b) can be used to identify and measure the TI of decametre-scale surface features which were not resolved in the original data.

**Acknowledgements**

This research is supported by an Australian Government Research Training Program (RTP) Scholarship.